\documentclass[superscriptaddress,secnumarabic,amssymb,amsmath,nobibnotes,showkeys,showpacs,aps,prd,nofootinbib,twocolumn]{revtex4}
\usepackage{color}
\usepackage{graphicx}
\usepackage{dcolumn}
\usepackage{bm}

\newcommand{\beq}{\begin{equation}}
\newcommand{\eeq}{\end{equation}}
\newcommand{\bey}{\begin{eqnarray}}
\newcommand{\eey}{\end{eqnarray}}

\begin{document}

\title{The Finslerian quantum cosmology}

\author{S S  De}
\email{ desatya06@gmail.com} \affiliation{Department of
Applied Mathematics, University of Calcutta, Kolkata 700009,
India}

\author{Farook Rahaman}
\email{rahaman@associates.iucaa.in} \affiliation{Department of
Mathematics, Jadavpur University, Kolkata 700032, West Bengal,
India}

\author{Nupur Paul}
\email{nnupurpaul@gmail.com} \affiliation{Department of
Mathematics, Jadavpur University, Kolkata 700032, West Bengal,
India}

\date{\today}

\begin{abstract}
We present a Friedmann-Robertson-Walker (FRW)  quantum cosmological model within the framework of Finslerian geometry. In this work, we consider a specific fluid. We obtain the corresponding Wheeler-DeWitt equation as the usual constraint equation as well as Schr\"{o}dinger equation following Dirac, although the approaches yields the same equation of time independent equation for the wave function  of the Universe. We provide exact classical and quantum mechanical solutions. We use the eigenfunctions to  study the time evolution of the expectation value of the scale factor. Finally we discuss the physical meaning of the results.

\end{abstract}

\pacs{04.40.Nr, 04.20.Jb, 04.20.Dw}

\maketitle

\section{Introduction}

It is argued that  there is a quantum mechanical contextual behind the classically detected Universe on large scales \cite{lapchinskii1977quantum}. One has to use quantum cosmology to describe quantum gravitational effects during the phase transition  in the very early stages of the Universe. Quantum cosmology is one of most significant emergent obtainment in the field of astronomy to modify the quantum ethos of the structure and evolution of the Universe. To explore cosmological phenomena and the evolution and nature of the Universe we mostly use Einstein's theory of relativity, which describes the notion of the Universe as a whole. Although the Einstein's field equation is most momentous revolutionary discovery that rules the field of astronomy, however, it fails to develop/describe some certain astrophysical phenomena specially the initial singularity \cite{pedram2008quantum}. If we could trace a possible resource to reach the beginning of the Universe from the prior echelon of big-bang, the concept of time and space should null and void which contradict our conventional wisdom. The idea, Quantum cosmology, involved as an alternative theory to develop the initial singularity that occurs from Einstein's field equation. It is quite significant attempt to assemble the theory of relativity and quantum gravity. The Wheeler-DeWitt equation conducts a weighty role in quantum gravity to obtain the wave function of the Universe which narrates the nature of Universe \cite{pedram2007schrodinger}. In quantum gravity, the quantum state of the Universe is ultimately described by the wave function which satisfies Wheeler-DeWitt equation. Wheeler-DeWitt equation is nothing but a mathematical expression which provides total energy of the Universe is
 zero, that is a possibility of a closed Universe bounded by space and time.

Recently, during the last two or three decades Finslerian gravity is an alternative gravitational theory based on Finsler spacetime, which is an anisotropic spacetime naturally arisen from a geometrical point of view. In fact, the motivation behind this approach lies on the observational facts, such as, anisotropy of microwave cosmic radiation, the flat rotation curve of spiral galaxies, accelerating expansion of the Universe and resolving it without dark energy etc {\cite{Sanders1996,Stavrinos2004,Vacaru2005,Chang2008,Stavrinos2013,Basilakos2013,Kouretsis2014,Chang2018}.} Present Finsler spacetime which is the FRW spacetime with Finslerian perturbation is, in fact, a~$(\alpha, \beta)$~-Finsler space \cite{paul2018cosmological}. Finslerian gravity in this spacetime can be regarded as an alternative doctrine of Einstein general relativity but it also includes it as a particular instance. Here we find out the solutions of classical gravitational field equations with the modified  barotropic equation of state $P=\omega\rho$, in the modified anisotropic pressure P of fluid distribution for different parameter values of $\omega$. These solutions include nonsingular accelerating Universes, as well as oscillatory Universe. We then construct the Hamilton for this system, so that this Hamiltonian or the corresponding Lagrangian can reproduce the classical gravitational equation as the Euler-Lagranges equation. On quantization, the canonical variables and the Hamiltonian operation $\widehat{H}$ are obtained. Then, the usual theory is employed in constructing the Wheeler-DeWitt equation. Similar technique has been used by Mebbarki \cite{mebarki2012towards} in the case of Randers-Finsler spacetime. The constraint equation $\widehat{H}\psi=0$ is imposed, and this, in fact, indicates that the Hamiltonian operator ~$\widehat{H}$~ annihilate the wave function of the Universe. Consequently, we find the Wheeler-DeWitt equation   approximation in a single gravitation degree of freedom which is the scale factor of FRW Universe. The equation is actually a time-independent Schr\"{o}dinger equation for the wave function of the Universe in many-world interpretation. Here also, following Dirac \cite{dirac1958theory}, Wheeler-DeWitt equation  has been constructed by imposing the constraint equation as an operator equation on the wave function. The resulting equation is found to be a time-dependent Schr\"{o}dinger equation for the wave function explicitly depend on time apart from its dependence on the scale factor. From this equation the same time-independent Schr\"{o}dinger equation (Wheeler-DeWitt equation) follows. We find the solutions for different parameter values of $\omega$. As in the case of Alvarenga el al \cite{alvarenga1998dynamical} the boundary condition might be imposed, and the inner product of any two functions can be defined to find out the expectation values of the scale factor. For a particular value of~$\omega$~, we have found this expectation value of the scale factor which represents a nonsingular Universe and accounts for the accelerated expansion of its.

{The Finsler spacetime violates the lorentz symmetry, in the article \cite{Pfeifer2012a} Pfeifer et al. extends the Einstein's gravity by generalizes the lorentzian type manifold where the dynamics of Finsler space is defined by action integral on the unit tangent bundle and also Stavrinos et al. \cite{Stavrinos2014} studies $f(R,T,F)$ theory in the context of Finsler space. Finsler perturbation method has been discussed in \cite{Xin2014} for the $\Lambda$CDM model. G. Papagiammopoulas et al. \cite{Papag2017} investigates about dynamical analysis and the growth index of linear matter perturbations of the Finsler-Randers cosmological model. The definite Lagrangian for Finsler-Rander space has been investigated in \cite{Stavrinos2012} in the presence of electromagnetic field.}

In this work we provide a quantum mechanical description of a FRW model with a specific fluid distribution in order to retrieve explicit mathematical expressions for the different quantum mechanical states.

The paper is organized as follows. In Sec. II, we construct the classical gravitational equation in Finsler spacetime with specific fluid structure. In Sec. III, the solutions have been found in different parameter values of $\omega$. In Sec. IV, Wheeler-DeWitt equation has been constructed from the Hamiltonian of the system. Solutions have been found there, and for a particular value of $\omega=-\frac{2}{3}$, the expectation value of the scale factor obtained. This expectation value of the scale factor has been compared with the scale factor of the corresponding classical case which also represents a nonsingular Universe with accelerated expansion of its. Finally, some concluding remarks have been made in the next section.

\section{Gravitational equations in Finsler space-time}

To describe the dynamics of the Universe, we consider the Finsler metric of the form as in \cite{li2014exact}
\begin{equation}\label{metric1}
F^2=y^ty^t-a^2(t)y^ry^r-r^2a^2(t)\bar{F}^2(\theta,\phi,y^\theta, y^\phi).
\end{equation}

This metric, in fact, has been shown there to be a $(\alpha, \beta)$-Finster space and describes FLRW spacetime with Finslerian perturbation.

This choice is motivated  by the  well known fact that at large scale our Universe is governed by the flat Friedmann-Robertson-Walker (FRW) line element. For constructing  a cosmological model, we assume   $\bar{F}^2$  as a quadratic in $ y^\theta$ and $y^\phi $.

The gravitational field equation in Finsler space is controlled by the base manifold of the Finsler space and the fiber coordinates $y^i$. The base manifold of the Finsler space controls the gravitational field equation in Finsler space and the fiber coordinates $y^i$ act as the velocities in the energy momentum tensor i.e. the velocities of the cosmic components. As a result, one can obtain the gravitational field equations in Finsler space  from the  metric (\ref{metric1}). We have assumed that for two dimensional Finsler space $\bar{F}$, $\bar{R}\mbox{ic}=\lambda$ i.e. the $\bar{F}$ is a constant flag curvature space. Note that this flag curvature of Finsler space generalizes the sectional curvature of Riemannian space.

We consider the general energy-momentum tensor for the matter distribution as
\begin{equation}\label{em-tensor}
T^\mu_\nu =(\rho +p_t)u^\mu u_\nu -p_tg^\mu_\nu+(p_r-p_t)\eta^\mu \eta_\nu,
\end{equation}
where $u^\mu u_\mu = -\eta^\mu \eta_\mu = 1$, $p_r$, $p_t$ are respectively the radial and transverse pressure for the  anisotropic fluid. The modified gravitational field equations in Finsler space-time are given by \cite{paul2018cosmological}
\begin{eqnarray}
8\pi_FG\rho&=&\frac{3\dot{a^2}}{a^2}+\frac{\lambda-1}{r^2a^2},\label{1e}\\
8\pi_FGp_r&=&-\frac{2\ddot{a}}{a}-\frac{\dot{a^2}}{a^2}-\frac{\lambda-1}{r^2a^2},\label{2e}\\
8\pi_FGp_t&=&-\frac{2\ddot{a}}{a}-\frac{\dot{a^2}}{a^2}.\label{3e}
\end{eqnarray}

We consider the modified  equation of state
\begin{equation}
P=\omega\rho,\label{4e}
\end{equation}
where $\omega$  is the modified equation of state parameter and $P$ is given by
\begin{equation}
P=(1+\omega)p_t-\omega p_r-K\frac{r^3}{2}F_a.
\end{equation}

Here the constant $K$ has the dimension of $(mass)^2,$ where $K\sim\frac{1}{L^2}\sim M^2.$

Therefore we can  write $K=m^2 ~~as~~  m\sim M.$

Hence
\begin{equation}
P=(1+\omega)p_t-\omega p_r-m^2\frac{r^3}{2}F_a.
\end{equation}

Note that here the anisotropic force is ~$\widehat{F} =\frac{r^3F_a}{2}$ where $F_a=\frac{2(p_t-p_r)}{r}$ and $m$ has the dimension of mass or inverse length (in natural unit $c=\hbar=1$ and the dimension of $\frac{m^2r^3}{2}F_a=m^2\widehat{F_a} \sim \frac{force}{area}$= pressure).

\section{Classical solutions}

With the gravitational field equations of Finsler spacetime (\ref{1e})-(\ref{3e}) and the equation of state (\ref{4e}), we have the following equation for the scale factor $a(t):$
\begin{eqnarray}
&&\ddot{a}+\frac{1+3\omega}{2}\frac{\dot{a}^2}{a}-\frac{m^2(1-\lambda)}{2a}=0,\label{9e}\\
\mathrm{or,}&&(1+3\omega)\dot{a}^2-m^2(1-\lambda)=Aa^{-(1+3\omega)},\label{10e}
\end{eqnarray}
where, $A$ is a integrating constant ( See Appendix A).

Using equations (\ref{9e}) and (\ref{10e}) we obtain the following equation for the scale factor $a(t)$:
\begin{equation}
\ddot{a}+\frac{A}{2}a^{-(2+3\omega)}=0.\label{11e}
\end{equation}

For different values of $\omega$, we will discuss the three cases as follows:\\

\textbf{Case I:} $\omega=0$

For $\omega=0$ the equation(\ref{9e}) yields

\[\ddot{a}+\frac{\dot{a}^2}{2a}-\frac{m^2(1-\lambda)}{2a}=0.\]

\textbf{Subcase I(a):} $A=0$

The solution is $a(t)=\hat{A}t,$  where $\hat{A}=m\sqrt{1-\lambda}$. Thus
\begin{equation}
a(t)=m\sqrt{1-\lambda}t.
\end{equation}

\textbf{Subcase I(b):} $A\neq0$

For $A\neq0$, eq. (\ref{10e}) yields,
\[\int\frac{\sqrt{a}da}{\sqrt{A+m^2(1-\lambda)a}}=\pm t+C.\]

Solving this we get,\\
$\sqrt{m^2(1-\lambda)}\left[\sqrt{a\left(\frac{A}{m^2(1-\lambda)}+a\right)}-\frac{A}{m^2(1-\lambda)}ln\left|\frac{\frac{A}{m^2(1-\lambda)}
+\sqrt{a}}{\frac{A}{m^2(1-\lambda)}}\right|~\right]$
\begin{equation}=\pm t+C
\label{rho}
\end{equation}
from (\ref{rho}) we can conclude that, here we can not obtain any explicit form of $a(t)$.


\textbf{Case II:} $\omega=-2/3$

The equation (\ref{11e}) becomes $\ddot{a}+\frac{A}{2}=0.$

Hence solving above equation and using (\ref{10e}), finally, we get

\begin{equation}
a(t)=\frac{1}{2}Bt^2+Ct+\frac{C^2+m^2(1-\lambda)}{2B},
\end{equation}
where, $B=-\frac{A}{2}$ and C is integrating constant. Then $a(t)=0$ if

\[\frac{1}{2}Bt^2+Ct+\frac{C^2+m^2(1-\lambda)}{2B}=0.\]

Where, Discriminant $=C^2-[C^2+m^2(1-\lambda)]=-m^2(1-\lambda).$ and it is negative if $\lambda<1$.

Hence a nonsingular Universe is possible.

\textbf{Case III:} $\omega=-1$

Equation(\ref{9e})-(\ref{11e}) give rise to the following equations
\begin{eqnarray}
\ddot{a}-\frac{\dot{a}^2}{a}-\frac{m^2(1-\lambda)}{2a})=0,\nonumber\\
\ddot{a}+\frac{A}{2}a=0.\nonumber
\end{eqnarray}

\textbf{Subcase III(a):} $A<0$

For negative values of A with $k^2=-A/2$, the above equations turn to be
\begin{equation}
\ddot{a}=k^2a,
\end{equation}

\begin{equation}
\dot{a}^2=k^2a^2-\frac{m^2(1-\lambda)}{2}\label{16e}.
\end{equation}

Equation (\ref{16e}) yields the  general solution as,
\begin{equation}
a(t)=a(0)coshkt+\frac{\dot{a}(0)sinhkt}{k}.
\end{equation}

For the generic initial condition there is no relationship between $a(0)$ and $\dot{a}(0)$.

In late time $(t>>1/k)$
\begin{equation}
a(t)\sim\left[a(0)+\frac{\dot{a}(0)}{k}\right]e^{kt}\propto e^{kt}.
\end{equation}

Which represents an accelerating Universe and an arrow of time. Indeed, such a situation has been discussed by Padmanabhan \cite{padmanabhan2017we} in the context of an `inverted' oscillator equation which is time-reversal invariant.

\[\left[note: k^2=\frac{m^2(1-\lambda)+\dot{a}^2(0)}{a^2(0)}\right]\]

\textbf{Subcase III(b): $A>0$}

For positive value of A with $k^2=A/2$,

In this case, we have the following equations \[ \ddot{a}+k^2a=0,\]\[\dot{a}^2+\frac{m^2(1-\lambda)}{2a})+k^2a^2=0.\]\

Here we obtain the following solution
\begin{equation}
a(t)=a(0)coskt+\frac{\dot{a}(0)sinkt}{k}.
\end{equation}

Hence, we obtain an oscillatory Universe.

\section{Hamiltonian formalism and Wheeler-DeWitt equation}

In order to quantize the gravitational field equation in Finsler geometry we first obtain the Hamiltonian for the system. For this the Lagrangian in the present case is taken as
\begin{equation}
{\L}=\frac{1}{2}\dot{a}^2a^{1+3\omega}+\frac{1}{2}\frac{(1-\lambda)m^2}{1+3\omega}a^{1+3\omega}.\label{20e}
\end{equation}

The canonical momentum  associated with $a$ is given by
\begin{equation}
\Pi=\frac{\partial {\L}}{\partial\dot{a}}=\dot{a}a^{1+3\omega}.\label{21e}
\end{equation}

The Euler-Lagrangian equation $\left[\frac{d}{dt}\left(\frac{\partial {\L}}{\partial \dot{a}}\right)-\frac{\partial {\L}}{\partial a}=0 \right]$ yields
\begin{equation}
\ddot{a}+\frac{1+3\omega}{2a}\dot{a}^2-\frac{m^2(1-\lambda)}{2a}=0.\label{22e}
\end{equation}

This equation is the same as in (\ref{9e}), and thus, justifies the choice of the Lagrangian (\ref{20e}).

Now, the Hamiltonian is constructed from this Lagrangian is given by
\begin{eqnarray}
H&=&\Pi\dot{a}-{\L},\\ \nonumber
&=&\frac{1}{2}\dot{a}^2a^{1+3\omega}-\frac{1}{2}\frac{m^2(1-\lambda)}{1+3\omega}a^{1+3\omega}.\nonumber
\end{eqnarray}

Using (\ref{21e}) and (\ref{22e}), we finally obtain (for ~$1+3\omega~\neq~0$~)
\begin{equation}H=\frac{1}{2}\Pi^2 a^{-(1+3\omega)}-\frac{1}{2}\frac{m^2(1-\lambda)}{1+3\omega}a^{1+3\omega}.\end{equation}

Now the Wheeler-DeWitt quantization leads to
  \[ H\rightarrow\widehat{H} ~,~\Pi\rightarrow\widehat{\Pi}=-i\frac{\partial}{\partial a}.~~~~~~~~~~  (c=\hbar=1)\]

Hence the Hamiltonian operator $\widehat{H}$ is
\[\widehat{H}=\frac{1}{2}\widehat{\Pi}^2a^{-(1+3\omega)}-\frac{1}{2}\frac{m^2(1-\lambda)}{1+3\omega}a^{1+3\omega},\]
assumes the following form
\begin{equation}\widehat{H}=-\frac{1}{2}a^{-(1+3\omega)}\frac{\partial^2}{\partial a^2}-\frac{1}{2}\frac{m^2(1-\lambda)}{1+3\omega}a^{1+3\omega}.   \end{equation}

{
Wheeler-DeWitt equation $\widehat{H}\psi(a)=0$  is obtained as
\begin{equation}~~~~~~~\frac{\partial^2\psi(a)}{\partial a^2}+\frac{m^2(1-\lambda)}{1+3\omega}a^{2(1+3\omega)}\psi(a)=0 .~~~~~~\end{equation}

Alternatively, Wheeler-DeWitt equation   can be obtained following Dirac \cite{dirac1958theory}
by imposing the constraint equation on the wave function of the Universe as,
\begin{equation}
i\frac{\partial\psi}{\partial t}=\widehat{H}\psi.\label{26e}
\end{equation}

This is in fact, the Schrodinger equation for the wave function $\psi(a,t)$.

Let,
\begin{equation}
\psi(a,t)=e^{-iEt}\phi(a).\label{27e}
\end{equation}

Then from eqs. (\ref{26e}) and (\ref{27e}) we get Wheeler-DeWitt equation for $\phi(a)$ as
\begin{equation}
\frac{\partial^2\phi}{\partial a^2}+\frac{m^2(1-\lambda)}{1+3\omega}a^{2(1+3\omega)}\phi+2Ea^{1+3\omega}\phi=0.
\end{equation}

Wheeler-DeWitt equation   can be obtained from the following Dirac \cite{dirac1958theory,unruh1989time} by imposing the constraint equation as an operator equation on the wave function of the Universe as
\[i\frac{\partial \psi}{\partial T}=-\widehat{H}\psi(a,T).\]

Here, the dynamical variable $T(t)$, where t is the time, is introduced in the Lagrangian and the constraint equation has been obtained from the variation of action with respect to $T$. The equation is, in fact, the Schr\"{o}dinger equation for the wave function ~$\psi(a,t)$~ if $t=-T$ is regarded as the time coordinate. In fact, ~$p_t$~ which is conjugate to time $t=-T$.

Now since \[\widehat{H}=-\frac{1}{2}a^{-(1+3\omega)}\frac{\partial^2}{\partial a^2}-\frac{1}{2}\frac{(1-\lambda)m^2}{1+3\omega},\]
the Wheeler-DeWitt equation becomes
\[i\frac{\partial\psi (a,t)}{\partial t}=-\frac{1}{2}a^{-(1+3\omega)}\frac{\partial^2\psi(a,t)}{\partial a^2}-\frac{1}{2}\frac{(1-\lambda)m^2}{1+3\omega}a^{1+3\omega}\psi.\]\

Let \[\psi(a,t)=e^{-iEt}\phi(a).\]

Then we get time-independent Wheeler-DeWitt equation
for $\phi(a)$ as
\begin{equation}
\frac{\partial^2\phi(a)}{\partial a^2}+\frac{m^2(1-\lambda)}{1+3\omega}a^{2(1+3\omega)}\phi(a)+2Ea^{1+3\omega}\phi(a)=0.\label{28e}
\end{equation}
Now, returning to the Lagrangian, it is to be noted that we can always add a constant term, say E, to the Lagrangian leaving the dynamical system unchanged. That is, we can take the Lagrangian to be \[{\L}=\frac{1}{2}\dot{a}^2a^{1+3\omega}+\frac{1}{2}\frac{(1-\lambda)m^2}{1+3\omega}a^{1+3\omega}+E.\]\

The Hamiltonian now becomes \[H={\L}=\frac{1}{2}\Pi^2a^{-(1+3\omega)}-\frac{1}{2}\frac{(1-\lambda)m^2}{1+3\omega}a^{1+3\omega}-E.\]\

On quantization we have \[\widehat{H}=-\frac{1}{2}a^{-(1+3\omega)}\frac{\partial^2}{\partial a^2}-\frac{1}{2}\frac{(1-\lambda)m^2}{1+3\omega}a^{1+3\omega}-E,\]
and consequently Wheeler-DeWitt equation  is usually obtained as $\widehat{H}\psi(a)=0$
or, \[\frac{\partial^2\psi(a)}{\partial a^2}+\frac{m^2(1-\lambda)}{1+3\omega}a^{2(1+3\omega)}\psi(a)+2Ea^{1+3\omega}\psi(a)=0.\]\

This equation is identical with the time-independent Wheeler-DeWitt equation obtained from the time-dependent Wheeler-DeWitt equation. Thus, time is naturally introduced from the above time-dependent Wheeler-DeWitt equation, if we consider it is derived from the Schr\"{o}dinger equation for the wave function of the Universe in which the coordinate time explicitly introduced in the wave function.

Now we will discuss the three cases for different values of $\omega$.

\textbf{Case I:} $\omega=0$

For $\omega=0$, equation (\ref{28e}) takes the form
\begin{equation}\frac{\partial^2\phi(a)}{\partial a^2}+m^2(1-\lambda)a^2\phi+2Ea\phi=0\end{equation}

Putting,~$x=m^2(1-\lambda)a+E$,~~we have
\begin{equation}
\frac{\partial^2\phi}{\partial x^2}+\frac{(x^2-E^2)\phi}{m^6(1-\lambda)^3}=0.\label{30e}
\end{equation}

Set \[\frac{1}{m^6(1-\lambda)^3}=\eta^2(if\lambda<1)\]

\[~~~~~~~~~~~~~~~~~~~~=-\eta^2(if\lambda>1)\]
\begin{equation}
\frac{\partial^2\phi}{\partial x^2}+\eta^2(x^2-E^2)\phi=0,  if~~(\lambda<1)\label{31e}
\end{equation}

For $\lambda<1$ equation(\ref{31e}) does not admit the harmonic oscillator solution.

For $\lambda>1$  equation (\ref{30e}) becomes

\[\frac{\partial^2\phi}{\partial x^2}+\eta^2(E^2-x^2)\phi=0.\]

This is formally identical with the time-independent Schrodinger equation for a harmonic oscillator with unit mass and energy $\frac{1}{2}\eta^2E^2$ \cite{alvarenga2002quantum}.  Possible values of energy are $n+\frac{1}{2},~~~n=0,1,2,......$
Thus
\[\frac{1}{2}\eta^2E_n^2=n+\frac{1}{2},~~~E_n^2=\frac{2n+1}{\eta^2},~~~E_n=\sqrt{\frac{2n+1}{\eta^2}}.\]\

Hence
\[E_n=m^3(1-\lambda)\sqrt{(1-\lambda)(2n+1)},~~~n=0,1,2.. ... ~~~~~~(\lambda<1)\]\

Hence
\begin{equation}\phi_n(x)=H_n\left[\frac{x}{m^{\frac{3}{2}}(1-\lambda)^{\frac{3}{4}}}\right]e^\frac{x^2}{m^3(1-\lambda)^\frac{3}{2}}.\end{equation}\

Finally one can get
\begin{equation}\psi_n(x)=e^{iE_nt}\phi_n(x).\label{33e}
\end{equation}

Here
\begin{equation}\psi^*\psi=H_n^2\left[\frac{a}{m\sqrt{1-\lambda}}+\sqrt{2n+1}\right]e^{\left[\frac{a}{m\sqrt{1-\lambda}}+\sqrt{2n+1}\right]}.\end{equation}

\[\left[Note: x^2\eta=\left(\frac{a}{m\sqrt{1-\lambda}}+\sqrt{2n+1}\right)\right]\]

In this case the wave function seems as wormhole-like wave function discussed in \cite{hawking1990spectrum}.

\textbf{Case II:} $\omega=-2/3$

Now the equation (\ref{28e}) assumes the form
\begin{equation}
\frac{\partial^2\phi}{\partial a^2}+m^2(1-\lambda)a^{-2} \phi+2Ea^{-1}\phi=0.
\end{equation}

This equation yields the following solution as

\begin{figure*}[thbp]
\begin{center}
\begin{tabular}{rl}
\includegraphics[width=7.cm]{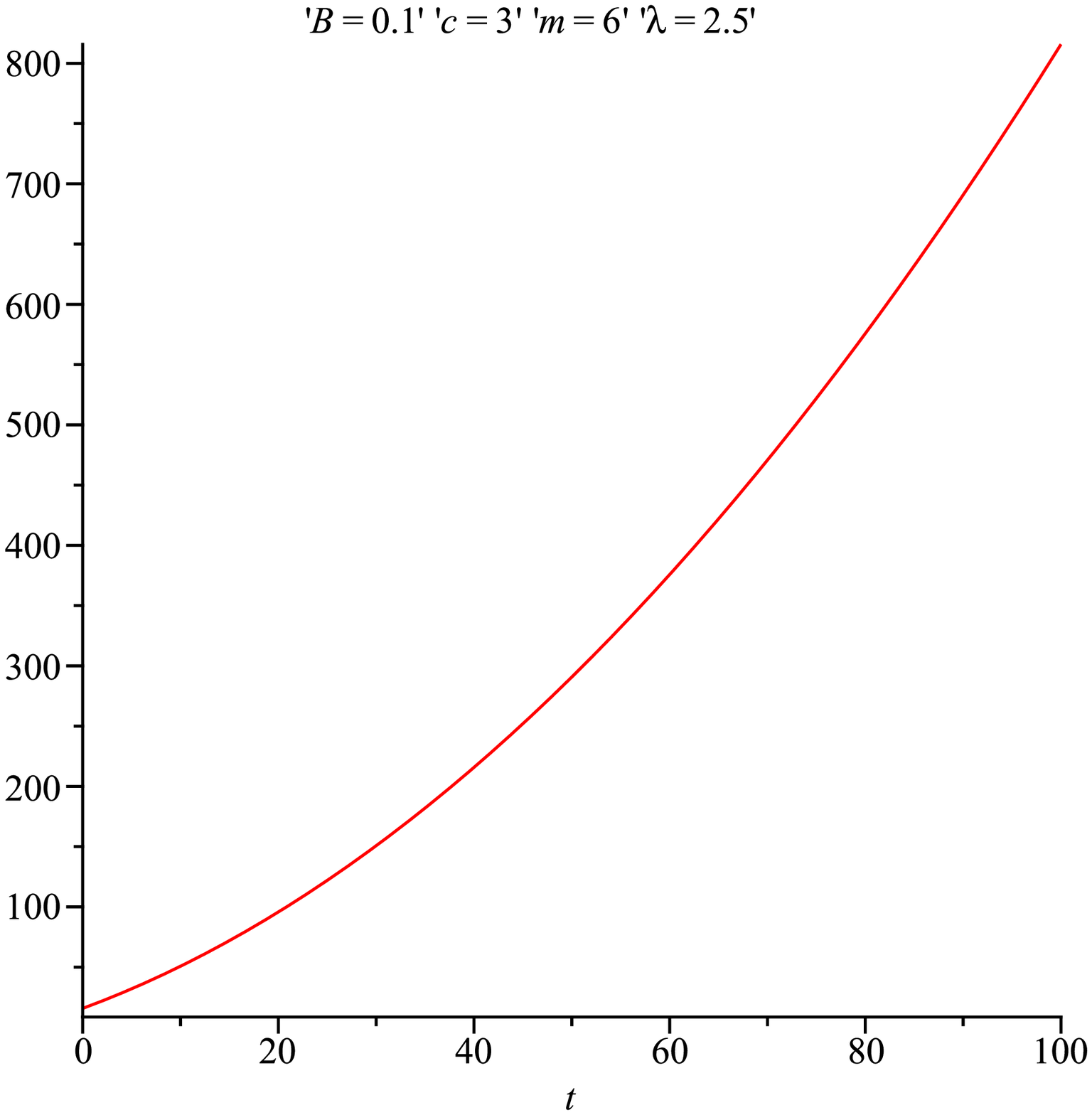}&
\includegraphics[width=7.cm]{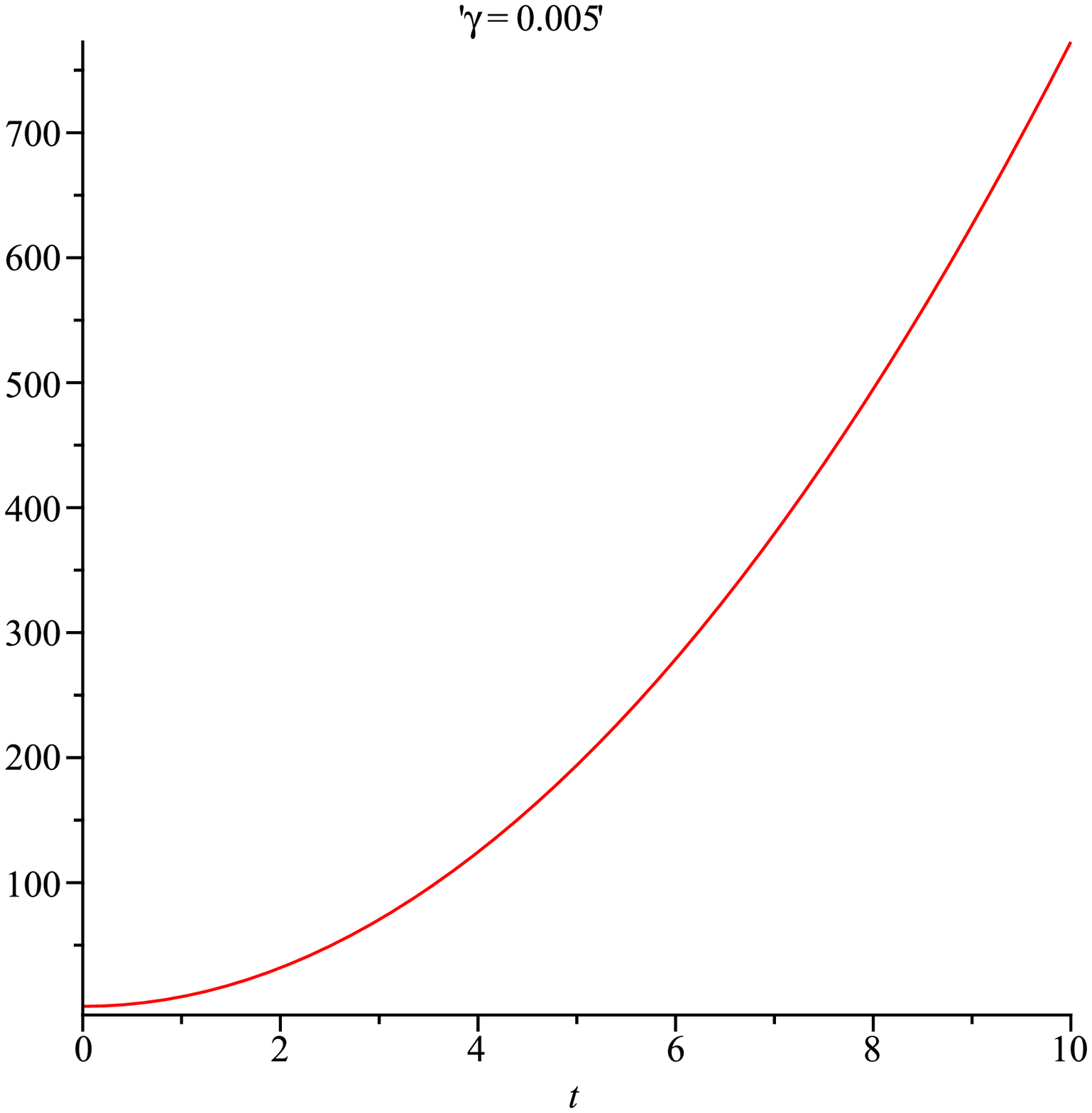}\\
\end{tabular}
\end{center}
\caption{Scale factor for classical model in the left panel and expectation value of the scale factor in the right panel  $\omega=-2/3$  .}\label{f6.1}
\end{figure*}

\begin{equation}~~~\phi(a)=\sqrt{a}\left[C_1J_{\nu}(z\sqrt{a})+C_2Y_{\nu}(z\sqrt{a})\right],~~~\end{equation}\
where
\[q=\sqrt{\frac{1}{4}+m^2(1-\lambda)},~~2q=\nu~~~and~~~z=2\sqrt{2E},\]\

We take the boundary conditions for the wave function ~~$\psi(a,t)$~~ as
\[\psi(0,t)=0 ~~and~~~\frac{\partial \psi(a,t)}{\partial a}|_{a=0}=0.\]

Taking the first boundary condition, we have $\phi(0)=0$ which implies
  $C_2=0,$ and consequently
\begin{equation}~~~~~~~\phi(a)=C_1\sqrt{a}J_{\nu}(z\sqrt{a}).~~~~~\end{equation}\

Hence the the wave function $\psi(a,t)$ takes the form as   \begin{equation}~~~ \psi(a,t)=C_1\sqrt{a}J_{\nu}(z\sqrt{a})e^{-iEt}~.~~\end{equation}~~

Following  Alvarenga, et.al \cite{alvarenga2002quantum}, the inner product of any two wave functions $\phi$  and  $\psi$  is taken as
\begin{equation}(\phi,\psi)=\int^\infty_0a^{1+3\omega}\phi^*\psi da~.\end{equation}

As the Hamiltonian operator $\hat{H}$ to be self-adjoint and also, since the solution (\ref{33e}) does not have finite norm, we have to construct the wave packet solution by superposition of the above solution in order to obtain wave function which can describe physical state. As, in Alvarenga, et.al \cite{alvarenga2002quantum} the superposition is taken as
\begin{equation}
\psi(a,t)=\int A(E)\psi_E(a,t)dE~,
\end{equation}
with a quasi-gaussian superposition factor and $\psi_E(a,t)$ as given in (\ref{33e}).

Then we have
\begin{equation}\psi(a,t)=\sqrt{a}\int^\infty_0z^{\nu+1}e^{-\gamma z^2-i\frac{z^2}{8}t}J_{\nu}(z\sqrt{a})dz~~[~as~ z=\sqrt{8E}],\end{equation}~~
where $\gamma$ is an arbitrary positive constant.

This integration can be found to be \cite{gradshteyn1980table}
\begin{equation}~~\psi(a,t)=\sqrt{a}~\frac{~e^{-{a^{\frac{1}{4B}}}}}{4B^2~},~~~~~~~~~~~~~~~~where,~~~ B=\gamma+\frac{it}{8}~~~~~~~~\end{equation}~~

With this, the expectation value of the scale factor with $\omega=-2/3$ can be found via
\[~~~~~~ ~~~<a(t)>=\frac{\int^\infty_0a^{1+3\omega}\psi^*(a,t)a\psi(a,t)da}{\int^\infty_0a^{1+3\omega}\psi^*(a,t)\psi(a,t)da},~~~~~\]
as
\begin{equation}~~~~~~<a(t)>\propto\left[\frac{t^2}{(72)^2\gamma^2}+1\right].~~~~\end{equation}

This represents a nonsingular bouncing Universe. It can be compared with the classical model for the case of $\omega=-2/3$ (see fig. \ref{f6.1}).\\

\textbf{Case III:} $\omega=-1$

For $\omega=-1$ equation (\ref{28e}) yields

\begin{equation}\frac{\partial^2\phi(a)}{\partial a^2}-\frac{m^2(1-\lambda)}{2}a^{-4}\phi+2Ea^{-2}\phi=0.\end{equation}

Solution can be found as

\[\phi(a)=\sqrt{a}\left[C_1J_{-\sqrt{\frac{1}{4}-2E}}\left(-\sqrt{\frac{-m^2(1-\lambda)}{2}}a^{-1}\right)\right]\]
\begin{equation}+\sqrt{a}\left[C_2Y_{-\sqrt{\frac{1}{4}-2E}}\left(-\sqrt{\frac{-m^2(1-\lambda)}{2}}a^{-1}\right)\right].\end{equation}

Due to this complicated solution, we cannot find the expected value of the scale factor, therefore, we would leave it for further discussion.

\section{Discussion and Conclusion }

This paper deals with  a quantum mechanical description of a FRW model in Finslerian background  with a specific fluid distribution in
order to  retrieve explicit  mathematical expressions for
the different quantum mechanical states.
In constructing Wheeler-DeWitt equation we have followed two different approaches - one the conventional and other following Dirac. We have shown that the two methods gave the same time independent Wheeler-DeWitt equation. Also, in Dirac method it has been shown that time has appears naturally. In minisuperspace gravitational degree of freedom is taken as to be scale factor of the Universe together with the parameter $\lambda$, a characteristic of the present Finsler space.  This $\lambda$ appears in Wheeler-DeWitt equation as the potential energy which disappears in Riemannian case, i.e. for $\lambda$ equals one. One can note that our consideration is a general one that includes Riemannian case. A rich variety of solutions are possible. But all solutions cannot be found for mathematical difficulties. Of course, in classical case we have obtained various solutions ranging from non-singular to oscillatory Universe. Also arrow of time has been obtained.

The time dependence of the scale factor in FRW Universe with a specific fluid was found by two different ways, namely, by solving Einstein-Finslerian equations and via Hamiltonian formalism. In our article we have discussed three cases for several values of equation of state parameter, $\omega$ in both classical and quantum approaches. Due to   complicated nature of the equations, we could not found explicit form of $a(t)$ and wave function in general case, but for the case $\omega=-\frac{2}{3}$,  we have found exact solutions in the context of Finsler gravity and Wheeler-DeWitt equation, which explain nature of the Universe.  By comparing both results for  $\omega =-2/3$, we could check the strength of this extrapolation.
  We have found the eigenfunctions and as a result acceptable wave packets were created by a suitable linear combination of these eigenfunctions. The time dependence of the expectation value of the scale factor has been found in connection of  the many-worlds interpretation of quantum cosmology. We have seen that the non-singularity is present in  FRW Universe. Moreover, this classical  Finslerian model predicts oscillatory and  accelerated Universe.
 Note that in the classical part for the case $\omega=-\frac{2}{3}$, if we substitute $\lambda$ by $1$, i.e. for Einstein gravity we find $a(t)=0$ for $t=-\frac{C}{B}$ (a constant), i.e. clearly big bang singularity occurs, which is a significant difference from our solutions on the Finsler manifold.
 Finally, one can conclude that this approach is physically pertinent and acceptable in order to contribute a new result in quantum Finslerian cosmology.

\appendix
\section{}
\label{a1}
We have
\begin{eqnarray}
&&\ddot{a}+\frac{1+3\omega}{2}\frac{\dot{a}^2}{a}-\frac{m^2(1-\lambda)}{2a}=0,\\
\mathrm{or,} &&d(\dot{a}^2)=[m^2(1-\lambda)-(1+3\omega)\dot{a}^2]d(lna),\\
\mathrm{or,} &&\frac{d(\dot{a}^2)}{(1+3\omega)\dot{a}^2-m^2(1-\lambda)}=-d(lna).
\end{eqnarray}

Finally we get,
\begin{equation}
(1+3\omega)\dot{a}^2-m^2(1-\lambda)=Aa^{-(1+3\omega)}.
\end{equation}

\section*{Acknowledgments}

FR would like to thank the authorities of the Inter-University Centre for Astronomy and Astrophysics, Pune, India for providing research facilities.   FR is also grateful to DST-SERB,  Govt. of India  and  Jadavpur University for financial support under RUSA 2.0.

\end{document}